\documentclass[12pt,preprint]{aastex}
\slugcomment{To appear in Astrophysical Journal}
\shorttitle{Effect of Microlensing on Transits}

\begin{document}

\title{Near-Field Microlensing and Its Effects on Stellar Transit 
Observations by {\em Kepler} }

\author{Kailash C. Sahu \& Ronald L. Gilliland}

\affil{Space Telescope Science Institute, Baltimore, MD 21218}

\email{ksahu@stsci.edu, gillil@stsci.edu}

\begin{abstract} 

In this paper, we explore the astrophysical implications of near-field
microlensing and its effects on stellar transit observations, with a
special emphasis on the {\it Kepler} mission. {\it Kepler} is a
NASA-approved mission whose goal is to detect a large number of
extrasolar, earth-like planets by obtaining near-continuous photometry
of $> 100,000$ F, G, and K dwarfs for four years. The expected
photometric precision of {\it Kepler} is 90 $\mu$mag (achieved in 15
minute samples), at which the effect of microlensing by a transiting
companion can be significant.  For example, for a solar-type primary
transited by a white-dwarf secondary, the maximum depth of the transit
is 0.01\%, which is almost entirely compensated by the microlensing
amplification when the white dwarf is at $\sim$0.05 AU.  The combined
effect of microlensing and transit increases to a net amplification of
150 $\mu$mag at an orbital separation of 0.1 AU, and 2.4 millimag at an
orbital separation of 1 AU.  Thus, the effect of microlensing can be
used to break the degeneracy between a planetary-mass object for which
the microlensing effect is negligible, and a more massive object of the
same size. For brown dwarfs at orbital separations of a few AU, the
effect of microlensing is several percent of the transit depth, and
hence the microlensing effect must be taken into account in deriving
the physical parameters of the brown dwarf.  The microlensing signal
caused by a neutron star or a black hole in a binary can be several
millimag, far exceeding the transit depth, and potentially detectable
even from ground-based observations. {\em Kepler} will be sensitive to
white dwarfs, neutron stars, and black holes in binaries through their
microlensing signatures.  These observations can be used to derive the
frequency of such compact objects in binaries, and to determine their
masses. 

\end{abstract} 

\section{Introduction} 

In a transit experiment where the goal is to look for obscuration of
light caused by the transiting body, the effect of microlensing by the
transiting body is generally thought to be negligible.  The main reason
is the following:  for orbital separations of a few AU, the Einstein
ring due to a stellar companion (the lens) is typically much smaller
than the size of the primary (the source), so that the maximum
amplification caused by the lens is small.  In addition, the lens has a
finite size, which may cause one or more of the rays to be occulted,
which makes the effect of microlensing small compared to the
obscuration caused by the lens.  In principle however, the effect of
microlensing can be important, and this effect becomes more important
when the binary companion is a degenerate star (i.e.  massive but
small) or if the orbital separation is large.  Gould (1995)  argued
that self-lensing in binaries in which at least one member is an
ordinary (non-compact) star (assuming $\gtrsim$ 0.3 mag amplification) 
would be exceedingly rare.  Marsh (2001) argued for better prospects in
finding microlensing signatures in main-sequence, white dwarf binaries,
but also concluded that  finding microlensing in binaries containing
neutron stars or black holes is unlikely.  At high photometric
precision, however, microlensing effects at {\em much} lower amplitude
can be important, and this effect must be  taken into account in
interpreting the observed light curves.

The Discovery-class mission, {\em Kepler: A Search for Terrestrial
Planets}, with an expected launch  in 2007, is a high-precision
photometric mission whose main goal is   to detect  a statistically
large number of extrasolar, earth-like planets, if such planets are
common around stars in the extended solar-neighborhood. {\em Kepler}
will provide such detections by obtaining extensive (four years),
near-continuous photometry on the same set of over 100,000 stars at
typical transit-depth precisions of 90 parts per million on the 15
minute sampling rate for $V \sim 12$ stars. An Earth-analogue transit
will typically last $\sim$10 hours and be  detected at $\sim$6 $\sigma$
per event which repeats once per year. (More details on {\em Kepler},
which is a NASA-approved discovery-class mission,  can be found at
http://www.kepler.arc.nasa.gov/). At such precision, the effect of
microlensing by a transiting compact and massive body can often be
significant.  For example, when the lens is a compact object such as a
white dwarf or a neutron star, the effect of microlensing can far
exceed the transit signal, causing a net positive amplification.  The
microlensing contribution due to a brown dwarf at 1 AU would be several
percent of the transit depth.  In this paper, we explore the
astrophysical implications of such microlensing in the  near-field
limit of a lens orbiting the source, with particular emphasis on the
upcoming {\em Kepler} mission. 

\section{Microlensing vs. Transit}

Gravitational microlensing studies over the past decade have revealed 
large numbers of events toward the Magellanic Clouds and the Galactic
Bulge  \citep{alcock00,udalski00,lassere01}. With near real-time
analysis of the photometry, many of the microlensing events are
detected early enough to provide alerts allowing for close and intense
follow-up monitoring of potentially interesting events. Such intense
monitoring has been carried out for several purposes such as, to look
for planets associated with the  lens (Albrow et al. 2001a; Gaudi et al.
2002); to determine limb-darkening parameters of the source (Albrow et
al. 2001b); to measure lens masses through parallax effects and orbital
motions of binary lenses (An et al. 2002); to determine  the caustic
crossing time-scales to determine the lens location (Afonso et al.
2000), etc.  The literature  on microlensing both from observational
and theoretical perspectives is now quite large and mature (e.g.  
Schneider et al. 1993, Paczy\'nski, 1996 \& Gould, 2001).  
 
By contrast, the much simpler effect of geometric obscuration resulting
in transits as one object passes in between a source and the observer
has received much less attention relative to microlensing (unless one
includes binary star eclipses in this context), although there has been
a surge of interest on this topic recently. In the case of classical
microlensing the events are rare, thus the  successful projects monitor
very dense stellar fields with large format detectors. However, the
microlensing signal can be quite large, so these projects require only
modest precision.  Transits of planets generally provide much smaller
signals: Jupiter transiting the Sun, for example, would yield an event
of 1\% depth. Furthermore, the inclinations are expected to be random,
so for planets with orbital  radii of $\gtrsim$1 AU, $>$99\% of
the existing systems would not exhibit transits  due to the small
chance of having the required alignment. For  terrestrial planets the
transit depth is even smaller: about one part in 10,000. Hence,
although seeking evidence of Earth-like, extrasolar planets (i.e., 
having a size comparable to the Earth at a distance of about an AU from
a solar-analogue host star) via transits has long been advocated as a
means to search for terrestrial planets (Borucki and Summers 1984),
this is a challenging  experiment.  With the stable observing
conditions provided by observations from space and the ability to
mosaic a significant number of CCDs it has become feasible to detect
true Earth-like extrasolar planets with a modest, dedicated mission
such as {\it Kepler}.

In a mission with such unprecedented photometric accuracy, it is
obviously important to make sure that the signature of an earth-like
planet is unique. Since a white dwarf has a radius comparable to that
of the Earth, a white dwarf in a one year orbit transiting a solar-like
star could in principle mimic the signature of a terrestrial planet
(although this can be easily eliminated in most cases by follow up
radial velocity measurements). This study is motivated by asking if the
photometric signal itself due to microlensing by the white dwarf, even
though it is only 1 AU from its host star, might be detectable.  Indeed
for the case  of a white dwarf in a one year orbit, the geometric
obscuration effect is at the level of one part in $10^4$, but the
microlensing effect is much larger, amounting to a few parts in $10^3$.
As we will show later, the light curve caused by the microlensing has a
form very similar to the transit light curve with a change of sign.
Taking into account the effect of gravitational microlensing, we will show
that white dwarfs transiting main sequence stars with orbital periods
of $\gtrsim$3 days would produce {\em positive} intensity signals, and
hence will not be a source of confusion with terrestrial planets for
{\em Kepler}. The purpose of this paper is to establish the framework
for calculating microlensing in this near-field regime, to establish
domains in source and lens characteristics where the effect is
measurable, and to explore whether the near-field microlensing signal
can be exploited to return astrophysical information such as the lens
mass.  In \S 3 we  discuss the amplification in the near-field 
regime and present a generalized formalism for generating light curves
taking the finite source size, the finite lens size, the inclination of
the orbit and the limb-darkening parameters into account, and discuss
how the simpler cases of point source etc. can be considered as special
cases of this generalized case.  In \S 4, we discuss the amplifications
caused by various types of degenerate and non-degenerate secondaries at
different orbital separations for different types of sources, and
discuss  the interesting cases in which the gravitational light
amplification and geometric obscuration effects are similar in
amplitude. \S 5 is devoted to a more detailed discussion of special
cases such as Earth-mass planets, Jupiters, brown dwarfs, white dwarfs,
neutron stars and black holes as secondaries in orbit around main sequence
stars.  We conclude in \S 6 with a summary of earlier sections and 
further discussion on the prospects of such observations with the
upcoming {\it Kepler} mission, and the possible astrophysical returns
from such observations.
 
{\section{Microlensing with Finite Source and Finite Lens}}

Some early theoretical work involving a finite source (and a point
lens) was carried out by Bontz (1979). A mathematical description
leading to an analytical solution for such a case can be found in
Schneider et al. (1992) and Witt and Mao (1994).  Dominik (1998) and
Gould \& Gaucherel (1997) have independently developed efficient
schemes for calculating the finite source effects using Green's theorem
and line integrals instead of following the conventional approach of
2-dimensional integration over the source. The numerical recipe 
developed by Sahu (1994 a,b) uses a polar coordinate system centered on
the source, which is particularly useful in considering
radially-symmetric effects such as the limb darkening, which is
followed here. 

The effect due to a finite {\it lens} has received much less attention;
nevertheless this has been discussed in some detail by  Bromley (1996)
and Marsh (2001), and more recently by Agol (2002) and Beskin and
Tuntsov (2002). The  literature dealing with the combined effect of
both a finite source and a finite lens is generally lacking.  Gould
(1995) considered the effect of microlensing in binaries and concluded
that only pairs of pulsars  would be interesting. However, his analysis
was done in the frame-work of the ``standard" microlensing where the
source lies within the Einstein ring radius of the lens, in which case
the amplification may be more than 34\%.  Marsh (2001) has generated
light curves for compact, degenerate binaries with small orbital
separations, where he assumed that the source is uniformly bright and
that the fainter of the two images is occulted by the lens. In order to
ensure that the results are valid for all cases observed by {\it
Kepler}, our calculations take appropriate limb-darkening parameters
into account, and do not assume {\it a priori} that the fainter image
is  occulted. 

In the following section, we discuss in detail the approach adopted
here in determining the effect of finite sizes of the source and the
lens simultaneously. Our light curve simulations are valid for low-mass
as well as  high-mass binaries at any orbital separation, and also
include an appropriate limb-darkening for the source and different
inclinations of the orbital plane. We note that in the extreme case
where the lens is a point lens, the net effect would be the same as
that of pure  microlensing. In the extreme case where the lens mass is
zero, the effect would be the same as that of a pure transit.  

\subsection{Formalism for Light Curve Simulation}

Let us first consider the source to be a point source. Let $D_L$ be the
distance to  the lens ($L$), $D_{LS}$ be the distance from the lens to
the source ($S$), and  $D_S$ be the distance from the observer to the
source as shown in Fig. 1.  Let $R_s$ represent the distance  between
the center of the lens and the source projected onto the lens plane. 

\begin{figure} \plotone{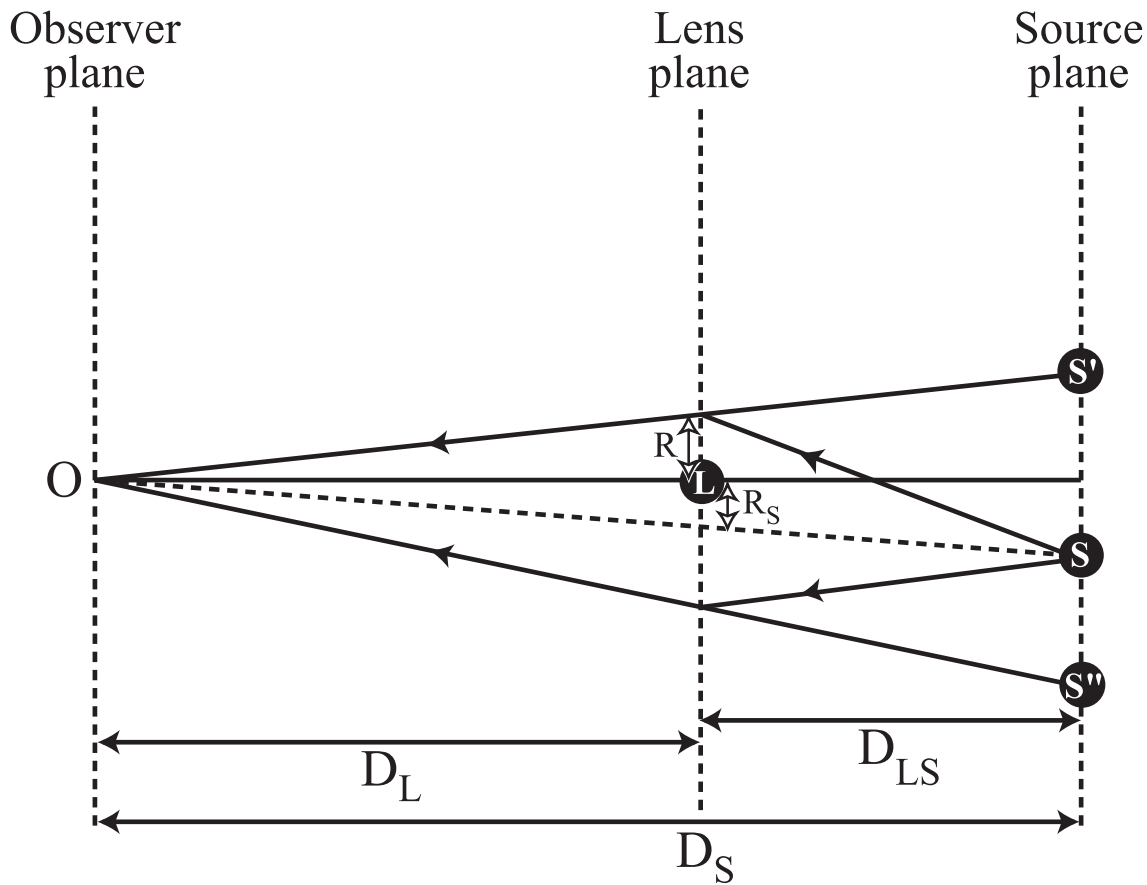} \caption{The geometry  of the
gravitational lensing is schematically shown here. The observer, the
lens and the source are located at positions O, L and S, respectively. 
D$_L$ is the distance to the lens, D$_S$ is the distance to the source,
and D$_{LS}$ is the distance from  the lens to the source. The lens (L)
produces two images of the source at positions S$^\prime$ and 
S$^{\prime\prime}$. At the lens plane, the lens, source and the two
images lie on a straight line. S$^\prime$ corresponds to the brighter
image which is formed outside the Einstein ring and is at least as
bright as the source itself.  S$^{\prime\prime}$ corresponds to the
fainter image which is formed inside the Einstein ring. } \end{figure}

The radius of the Einstein ring at the lens plane, $R_E$,
can be written as

\begin{equation} R_E^2 = {{4GM_LD}\over {c^2}}, \quad \quad D \equiv {D_{LS} D_L\over D_S}  \end{equation}
 
\noindent where $M_L$ is the the mass of the lens.

The lens produces two images of the source, $S^\prime$ and
$S^{\prime\prime}$, as shown in Fig. 1. At the lens plane, these two
images are situated on the line joining the source and the center of
the lens, one inside and the other outside the Einstein ring. The
distance between the lens and the two images can be expressed as (e.g.
see Paczy\'nski 1996 for details)

\begin{equation}R_{+,-} = 0.5 [R_s \pm (R_s^2 + 4R_E^2)^{1/2}] \end{equation}.

The amplification of these two images are given by

\begin{equation}A_p^{+,-} =  {{u^2+2}\over 2u(u^2+4)^{1/2}} \pm 0.5,
\quad \quad u \equiv R_s/R_E  \end{equation}.

Thus we see that the image outside the Einstein ring is the brighter
of the two images, and is at least as bright as the source
itself. The combined amplification $A_p$ for a point source is given by

\begin{equation}A_p =  A_p^+ + A_p^- = {{u^2+2}\over u(u^2+4)^{1/2}} \end{equation} 

If $u$ is replaced by  the minimum impact parameter $u_0$, we get the maximum 
amplification $A_{p}^0$.
 
Let us now discuss the case of an extended source, which can be treated
as an ensemble  of point sources. We will assume the lens to be
extended, opaque and spherically symmetric.  For an extended source, it
is convenient to work in the lens plane,  and  Fig. 2 schematically
shows such a case, where the source is larger than the Einstein ring
itself.  The two circles A and B represent two parts of the source,
whose images are (A$^\prime$, A$^{\prime\prime}$) and (B$^\prime$,
B$^{\prime\prime}$), respectively. The brighter images (A$^\prime$ and
B$^\prime$) are  outside the Einstein ring, none of which (A$^\prime$)
is occulted by the lens. The fainter images (A$^{\prime\prime}$,
B$^{\prime\prime}$) are inside the Einstein ring, one of which
(A$^{\prime\prime}$) is occulted by the lens. It is worth noting that
the images of all the parts of the source are displaced with respect to
the original positions such that no two images overlap. Microlensing
preserves the surface brightness, and any amplification caused by
microlensing is caused by a proportional increase in the size of the
image with respect to the original size.

\begin{figure} \plotone{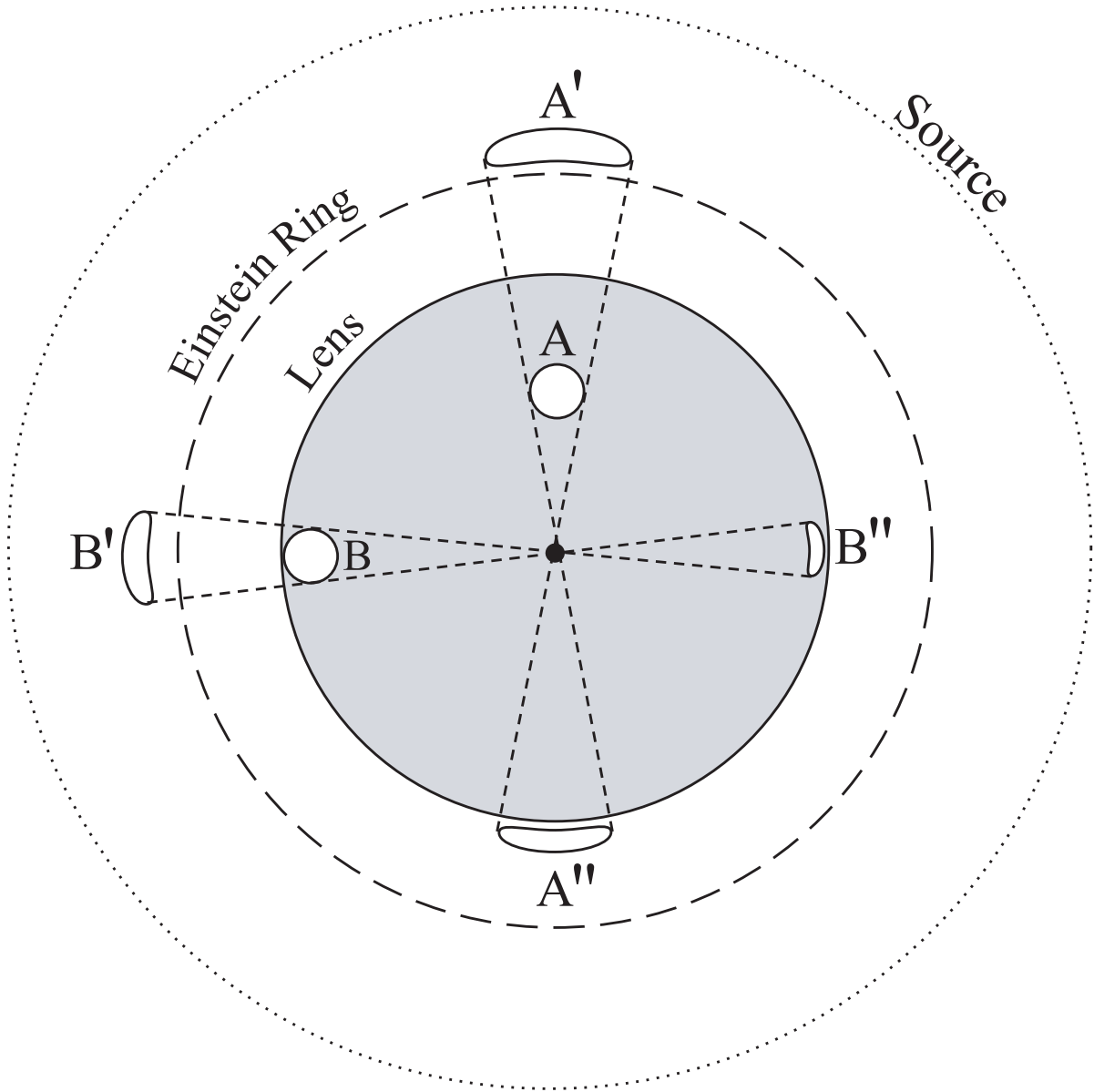} \caption{A schematic showing the
positions of the lens, the source and the images in the lens plane. The
shaded circle is the lens, and the dashed circle is the Einstein ring,
and the dotted circle is the source. The circle A represents a part of
the source whose images are A$^\prime$ and$A^{\prime\prime}$.  None of
these two images are occulted by the lens. The circle B represents a 
part of the source farther away from the lens, whose images are
B$^\prime$ and $B^{\prime\prime}$. Notice that the fainter image
corresponding to B$^{\prime\prime}$ is occulted by the lens, whereas
the brighter image corresponding to B$^{\prime}$ is not occulted. }
\end{figure}

The amplification of an extended source would then be given by
(equation 6.81 of Schneider et al. 1993) 

\begin{equation}A = {\int {d^2y \ I(y) \  A_p(y)} \over {\int d^2y \ I(y)}} \end{equation} 

\noindent where $I(y)$ is the surface brightness profile of the source,
$A_p(y)$ is the amplification of a point source at position $y$, and the
integration is carried out over the entire surface of the source.
 
For a lens with finite size, the above equation must be rewritten as 

\begin{equation}A = {\int d^2y \ I(y) \ [A_p^+ (y) + A_p^-(y)] \over \int d^2y \
I(y)} \end{equation} 

\noindent where the integration is carried out over all the points in
the image plane  which are not occulted by the lens.

Note that for a finite-size, uniformly bright source, the maximum
amplification (corresponding  to a point lens which is perfectly
aligned with the source) is given by

\begin{equation} A_{max} = [1 + {4R_E^2\over{r_S^2}}]^{1\over{2}} \end{equation}

\noindent where $r_S$ is the radius of the source scaled to the lens plane.

\begin{figure}
\plotone{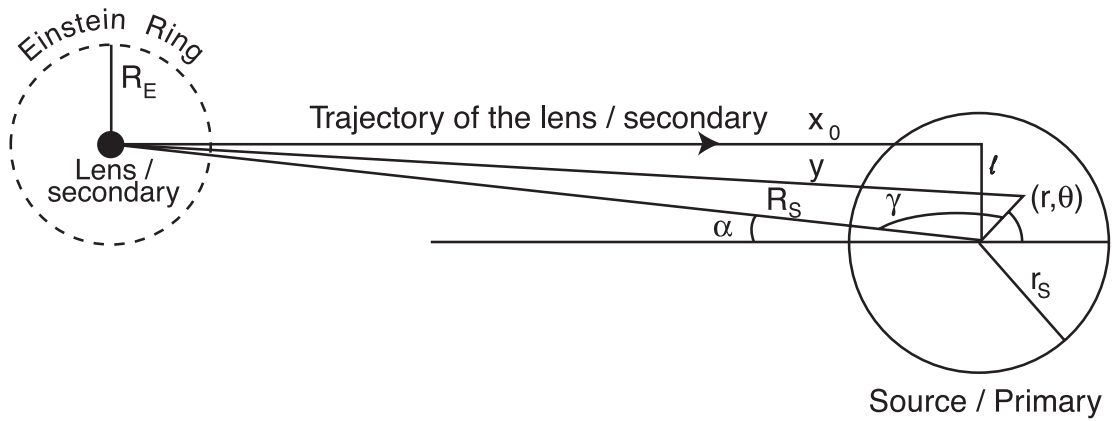}
\caption{The geometry used for calculating the microlensing and the
transit contributions, taking the effect of the extended source and the
extended lens into account. $l$ is the  minimum impact parameter and
$r_S$ is the source radius. A polar coordinate system is defined as
shown here to integrate over the source. The effect of the inclination
(not shown in the figure) is taken into account through appropriate
changes in the impact parameter. We take the effect of the transit and
the microlensing simultaneously into account by taking the deflections
and  amplifications of the unocculted rays from every point in the
source.  To check the correctness of the algorithm used, the 
contribution of a pure transit was calculated in two ways: by
specifying the mass of the lens to be zero, and by forcing the
deflections caused by the gravitational potential of the  lens to be
zero. Both gave identical results as expected.}
\end{figure}

In order to carry out the integration given by equation (6), let us follow
Fig. 3 where $R_s$ is the distance between the lens and the center of
the source projected onto the lens plane, and $l$ is the impact
parameter (the minimum distance between the center of the lens and the
center of the source at the lens plane as the lens traces the path
shown in the figure).  Let us choose a circular coordinate system and
let ($r,\theta$) be the representative point which is at a distance $y$
from the lens. From Fig. 3 we see that
 
\begin{equation}y = \sqrt{R_s^2 + r^2 - 2 \ R_s \ r \ cos(\gamma )}
\end{equation} 

\noindent where  

\begin{equation}R_s=\sqrt{x_0^2 + l^2},\ \gamma = \pi -\theta -\alpha \
and \  \alpha = sin^{-1}(l/R_s) \end{equation}

\noindent i.e.

\begin{equation}y = \sqrt{ R_s^2 +r^2 +2 \ R_s \ r \
cos(\alpha+\theta)} \end{equation} 

The two images of the point ($r,\theta$) are formed at distances $y_+$
and $y_-$ from the center of the lens along the line joining the lens
and the point ($r,\theta$) such that

\begin{equation}y_{+,-} = 0.5 [y \pm (y^2 + 4R_E^2)^{1/2}] \end{equation}.

The image is not occulted if
\begin{equation} y_{+,-} > r_l \end{equation}

where $r_l$ is the radius of the lens.

The amplification corresponding to these two images can be written as 

\begin{equation} A_{+,-}(y) =  {{{(y/R_E)}^2 + 2} \over { {2(y/ R_E)}
{[(y/R_E)^2}+4]^{1/2}}} \pm 0.5  \end{equation}

In calculating the combined amplification $A(y) = A_{+}(y) + A_{-}(y)$,
we consider only the images which satisfy the condition (12) to ensure
that the image is not occulted by the lens.

Using equation(6), the total amplification can be calculated by integrating
over the source using

\begin{equation} A = {{\int_0^{r_S} \int_0^{2\pi} {A(y)} \
I(y)  \  dr \ d\theta} \over { \int_0^{r_S} \int_0^{2\pi}  I(y) \ dr
\ d\theta}} \end{equation}

This is an expression suitable for numerical integration which will
take the finite size of both the lens and the source into account. The
light curve  can be reproduced as a function of time $t$, which is
related to $x_0$ through the relation $x_0 = (R_E/t_0)t$ where  $t_0$
is the timescale of microlensing (which is the time taken by the lens
to cross its own Einstein ring). $y$ in turn, is related to $x_0$
through equation 8 and 9.

Since we are mostly interested in the effect caused by the secondary in 
a binary, we need to include the effect of the inclination, which can be 
incorporated through a suitable choice of the minimum impact parameter. 
The light curves can then be calculated with the amplification as a 
function of the phase (see Fig. 3). 

\subsection{Determination of the Lens Mass}

We also provide here the equations that allow the mass of the lens to
be derived from photometry in the limit where microlensing
amplification dominates the light curve, as for example would be the 
case for a white dwarf orbiting a main-sequence star at 1 AU.  If
multiple, periodically repeating events have been observed, Kepler's
law provides a relation between the orbital period ($P$), semi-major
axis ($a$) and the sum of the masses:

\begin{equation}P^2 = {{4 \pi ^2 a^3 }\over {(M_S + M_L)G}}  \end{equation}

\noindent where $M_S$ is the mass of the
source which may be constrained from independent observations
(such as spectroscopy and spectral classification and/or with radial
velocity observations), and $M_L$ is the lens mass for which an
estimate is desired.

Assuming $D_{LS} \sim a < < D_L$, equation (7) can be rewritten as

\begin{equation}M_L = {{(A_{0}^2 - 1) r_S^2 c^2} \over {16 G a}}  \end{equation}

\noindent where $r_S$ is the  radius of the source star scaled to the
lens plane, and $c$ is the velocity of light. $A_{0}$ is the maximum
amplification during the microlensing event which  we have assumed to
be $\sim A_{max}$.  Note that for source sizes much larger than the
Einstein ring radius, the maximum amplification during the microlensing
event ($A_0$) is nearly equal to $A_{max}$ given by equation 7
regardless of the impact parameter, provided that the lens actually
transits the source. This is demonstrated by  the flat-topped nature of
the numerically-simulated lightcurves (Fig. 10, 14 and 15), and is
consistent with discussions in several other papers (e.g. Agol 2002).
Thus $A_0 \sim A_{max}$ is generally a good approximation. If
independent observations provide  constraints on the source radius,
then equation (16) can be used to  solve for orbital
separation, $a$, and $M_L$.  

An additional constraint comes from the duration of the transit ($T_t$) which,
for a circular orbit and 90$^\circ$ inclination, can be written as

\begin{equation} T_t = {{2 r_s P} \over {2 \pi a}} =
{{2 a^{1/2} r_S} \over {{[(M_L + M_S)G]^{1/2}}}} \end{equation}.

Combining equations (16) and (17) to eliminate $r_S$ 

\begin{equation} (A_0^2 -1) T_t^2 =  {{64 M_L a^2} \over
{(M_L + M_S) c^2}} \end{equation}

After a little algebra, this can be rewritten as

\begin{equation}M_L = {{M_S (A_0^2 - 1) T_t^2 c^2} \over
{64 a^2 - (A_0^2-1) T_t^2 c^2}} \end{equation}

This equation can be used to derive the mass of the lens if $M_S$ and
$a$ are known.

We can also combine equations (16) and (17) to eliminate $a$, and write

\begin{equation} {A_0^2 - 1} =  {{4 G^2 M_L (M_L + M_S) T_t^2} 
\over {c^2 r_S^4}} \end{equation}

This equation can be solved to express the lens mass

\begin{equation}M_L = {{{[G^2 M_S^2 T_t^2 + (A_0^2 -1) c^2 r_S^4]^{1/2}} -
G M_S T_t} \over {2 G T_t}} \end{equation}

This equation can be used to derive the mass of the lens if $M_S$ and
$r_S$ are known.

We will return to a discussion of lens mass determination with an
example in \S~5.2.

\subsection{Generalization to a limb-darkened source}

Since the polar coordinate system we have chosen is centered on the
source  (Fig. 3, equation 8), it is easy to incorporate limb-darkening for
the source which has an azimuthal symmetry. Recently,  the limb
darkening parameters were derived for HD 209458 using the high
signal-to-noise observations obtained with the Space Telescope Imaging
Spectrograph aboard  Hubble Space telescope in the wavelength region
582 to 638 nm (Brown et al. 2001). Based
on these observations, a limb-darkening of the form 
  \begin{equation} I(\mu) = 1.0 - u_1(1-\mu) - u_2(1-\mu)^2  \end{equation} 
 \noindent was used, where $\mu$ is the cosine of the angle between the
line  of sight and the normal to the local stellar surface (Claret \&
Gimenez 1990).  We have used $u_1 = 0.3$ and $u_2 = 0.35$ as applicable
to HD 209458 in the wavelength region 582 to 638 nm, in generating 
all the light curves here. We should note,
however, that the formalism  discussed in this paper can be used to
take any other specific form of the limb-darkening as well.

\subsection{Special case: point lens and zero-mass lens}

In the limiting case of a point lens, the results should be identical
to the results obtained from other standard treatments of microlensing
involving a finite source and a point lens. In  particular, equation
(7) can be directly used to calculate the amplification for the special
case where the lens and the source are  perfectly aligned. This
provides a direct test (although only for  the special case of a point
lens) which was used to check the algorithm discussed above.  The
previous results obtained by Sahu (1994b) and confirmed by Dominik \&
Hirshfeld (1996) also served as a test case for a point lens, to check
the  correctness of the simulated light curves. We have also checked
our calculations with a few special cases calculated by Marsh (2001)
and obtained consistent results.

In the limiting hypothetical case where the mass of the lens is zero
(but the  size is finite), the results obtained here should be
identical to the results obtained for pure transit. The transit light
curve simulated by  Brown et al (2001) for HD 209458 serves as a
test case for this purpose. Our algorithm was used to simulate a light
curve for HD 209458 using the same  limb-darkening parameters suggested
by Brown et al (2001). The resultant light curve was found to be
identical to the one published by Brown et al (2001) which 
provided a further check on the validity of the algorithm used here. 

The treatment for a zero-mass lens is  equivalent to assuming that the
images are not deflected, the amplification  corresponding to the
brighter image is one, and the amplification corresponding  to the
fainter image is zero. This, in turn, is equivalent to a restricted
version of the algorithm where all deflections and amplifications are
ignored.  The light curves were simulated using both these approaches,
and the resultant light curves were identical as expected. 

\section{Net amplifications for various secondaries}

The maximum net amplification caused by a secondary companion (after
taking  the finite sizes for the source and the lens into account)
generally  corresponds to the configuration where  the lens and the
source are perfectly aligned (except in a few special cases where the
maximum amplification occurs just before or after the egress as
explained later). This configuration corresponds  to $R_s = 0$ (equation 8),
which can be used in equations (8-14) to calculate the amplification. An
analytical solution for such a configuration was derived by Marsh
(2001).

The maximum amplification thus calculated for a variety of secondaries
(Jupiter, brown dwarf, white dwarf, neutron star and black hole) at
various orbital separations are shown in figures 2 to 7, assuming $D_L > >
D_{LS}$. Figs. 4-8
assume the source to be a solar type star with radius 7 $\times
10^{10}$cm. The radii and the masses assumed for the different
secondaries are given in Table 1. 

\begin{table*}
{\caption{Secondary (Lens) Characteristics}}
\begin{tabular}{lcccc}
\\
\hline
Secondary && Radius    & Mass   \\
         & &($R_\odot$) & ($M_\odot$) \\
\hline
Jupiter&& 0.1 & 0.001\\
Brown Dwarf & &0.1 & 0.05\\
White Dwarf&& 0.01& 0.6\\
Neutron Star&& $2.8 \times 10^{-5}$& 1.4\\
Black Hole & &$< 1\times 10^{-5}$&8.0\\ 

\hline
\end{tabular}
\end{table*}

In most cases that we consider, the secondary (lens) will provide an
amount of light in the system that is small compared to that from the
primary (source). We therefore do not provide explicit corrections to
light curves from  the luminosity of the lens itself.  We also note the
obvious, yet subtle, point that the light from the lens is constantly
present before, during and after an occultation/microlensing event. 
Therefore, during the  transit the only effect from the light from the
secondary is a dilution of the light curve changes in the same
proportion as the secondary  contributes to the total light light away
from the transit.  For a canonical case of a white dwarf orbiting a
solar-like star in which the surface brightness of the two components
are similar, the white dwarf will have a flux about $10^{-4}$ that of
the primary implying a 0.01\% dilution effect. ($Kepler$ might be able
to detect the secondary eclipse, and this would be interesting, but
correcting for the minor change of inferred microlensing amplitude is
not relevant). Exceptions to this rule can arise, e.g. when the primary
is intrinsically small such as a brown dwarf, the amplifications of
which are shown in Fig. 9. In the case of a white  dwarf lensing a
brown dwarf, this dilution effect could be a dominant effect which can
be applied in a trivial manner using the expected luminosities of the
two objects in the observational bandpass.

\begin{figure} 
\plotone{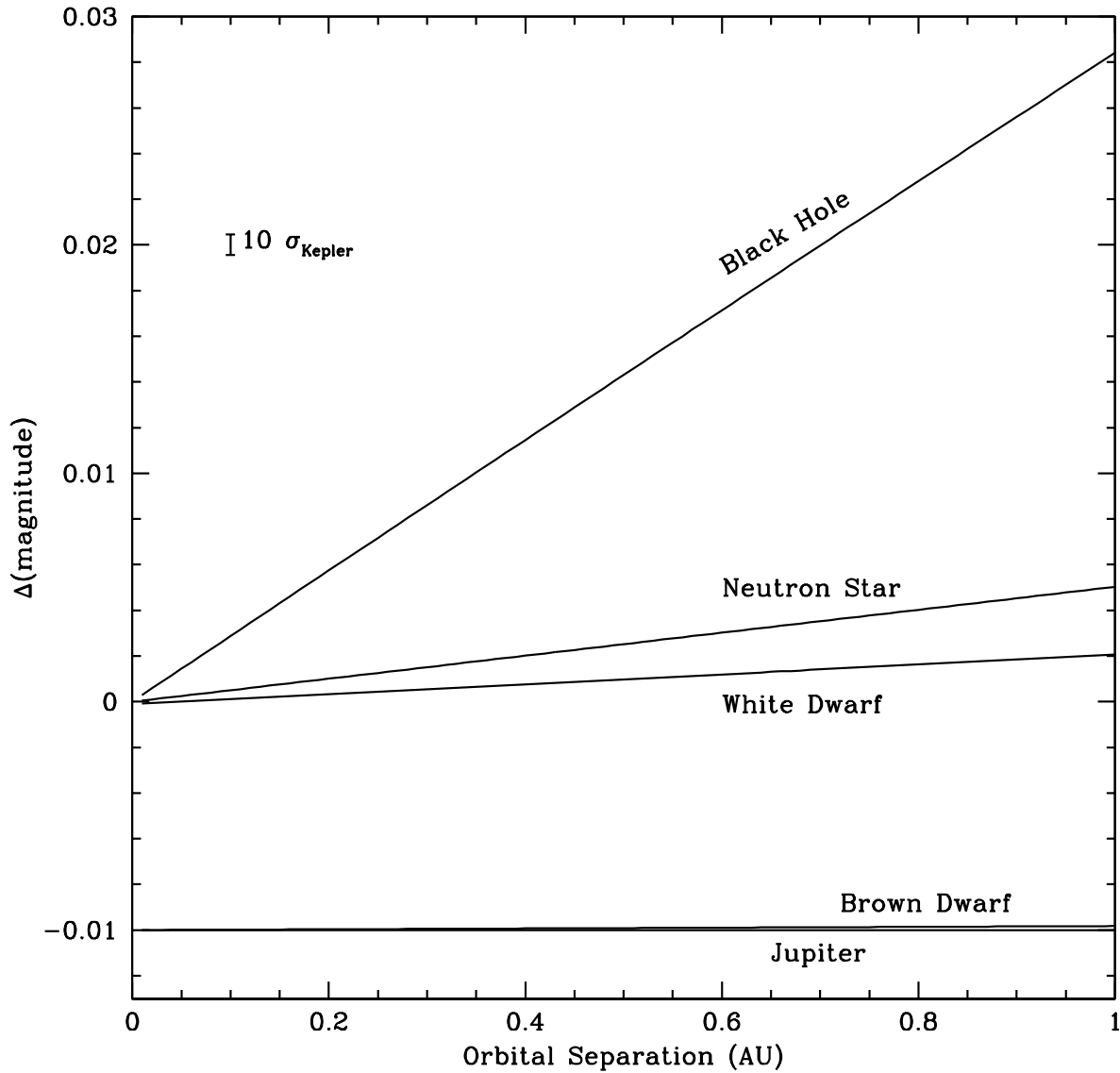} \caption{The net amplification
caused by different types of secondaries is plotted as a function of
orbital separation after taking both the  transit and the microlensing
contributions into account. See Table 1 for the details on the radii
and the masses used for the different types companions. The source is
assumed to be a solar type star, and the inclination is assumed to be
90$^\circ$. The errors expected from the {\em Kepler} mission are too
small to be shown in this scale, so the error bar shows 10 times
$\sigma_{Kepler}$ where $\sigma_{Kepler}$  refers to the expected
photometric error per 15-minute sample at $V=12$. Note that at 1 AU,
the amplification caused by a black hole secondary is $\sim$2.5\% which
is within reach of ground-based observations, and could be detectable
in ground-based survey programs such as OGLE III. } \end{figure}

Fig. 4 shows the effects of different secondaries for orbital
separations  of 0 to 1 AU and Fig. 5 shows the same for orbital
separations of 0 to 5 AU.  Note that the radii of a Jupiter and a brown
dwarf are identical. However, while the microlensing contribution due
to a Jupiter is negligible, the microlensing contribution due to a
brown dwarf is appreciable at Kepler's sensitivity even at 1 AU. As a
result,  the $\Delta$mag caused by these two types of secondaries are
different at Kepler's sensitivity at 1 AU, which is shown in Fig. 4.
This effect must be taken into account in deriving the radius of a
brown dwarf (more details on such a case are given later).
   
\begin{figure}
\plotone{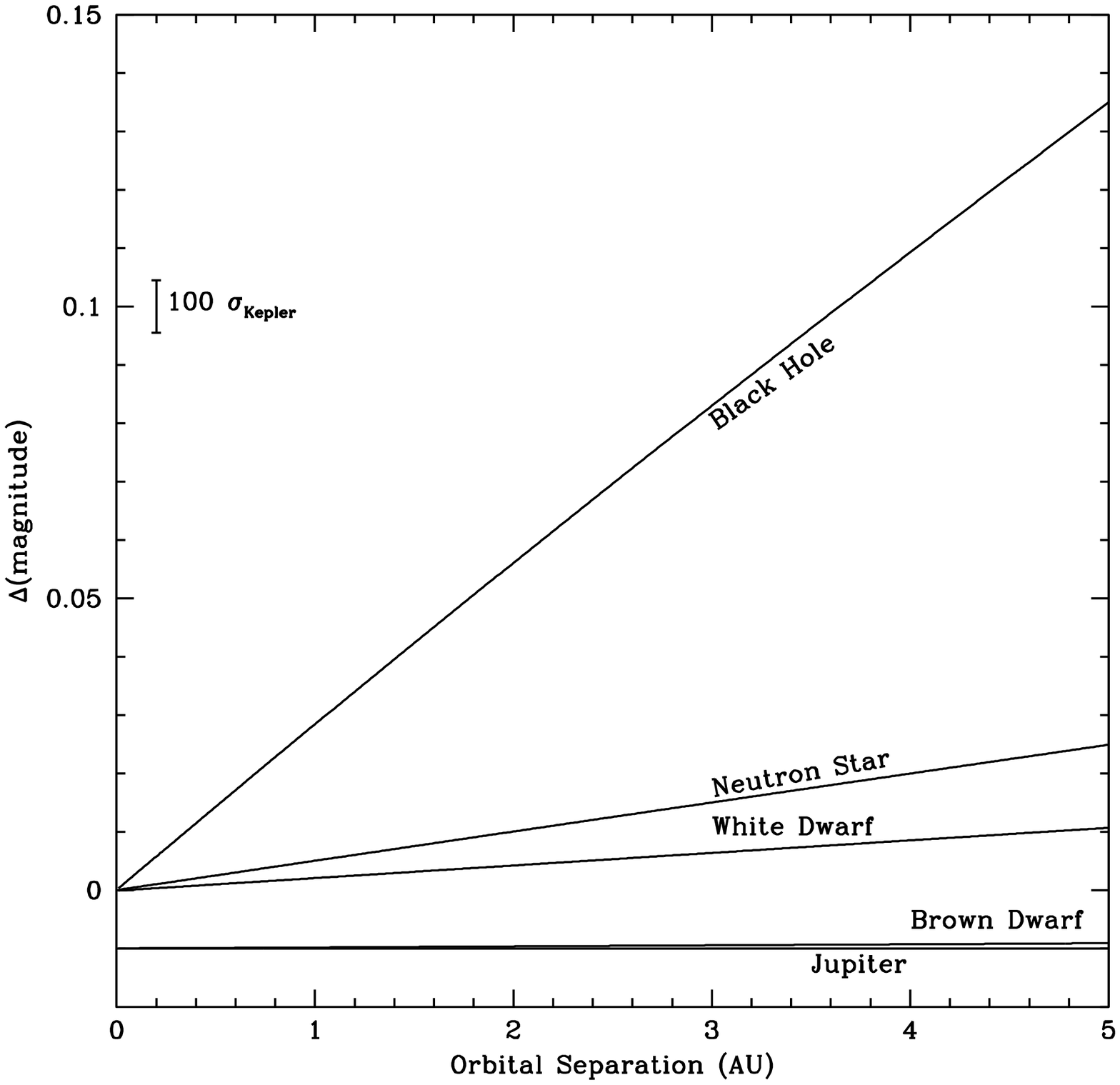}
\caption{Same as the previous figure but for a larger range of 
orbital separations (0 to 5 AU). The error bar shown here 
corresponds to 100 times $\sigma_{Kepler}$.
At an orbital separation of 5 AU, the
amplification caused be a white dwarf is more than 10 millimagnitudes.
The amplifications caused by neutron star and black hole binaries 
at 5 AU would be well within the reach of ground-based survey programs.}
\end{figure}

For white dwarfs, neutron stars and black holes, the net effect is
dominated by the microlensing at any orbital separation, and the
transit contribution is negligible (Figs. 4-5). In order to show the
{\em Kepler} sensitivity,  it had to be multiplied by factors of 10 and
100 in Figs. 4 and 5, respectively. The net effects caused by the
neutron star and the black hole at 1 AU,  5 and 30 millimag
respectively, are large enough in principle to be detectable even from
ground-based observations. 

An enlarged view of the effects of the brown dwarf and Jupiter are
shown in Fig. 6 and Fig. 7, for orbital separations of 0-5 AU and 0-100
AU, respectively. As seen from Fig. 6, the microlensing contribution
due to a brown dwarf is significant even for small orbital separations
and must be taken  into account in estimating its radius from the
observed light curves. Even for a Jupiter-mass secondary, the
microlensing contribution starts to become relevant  at orbital
separations of 5 AU or more at {\em Kepler}'s sensitivity. From Fig. 7
we see that the microlensing contribution from a brown dwarf almost
exactly cancels out the transit contribution at an orbital separation
of $\sim$55 AU, beyond which the net amplification is positive. 

\begin{figure}
\plotone{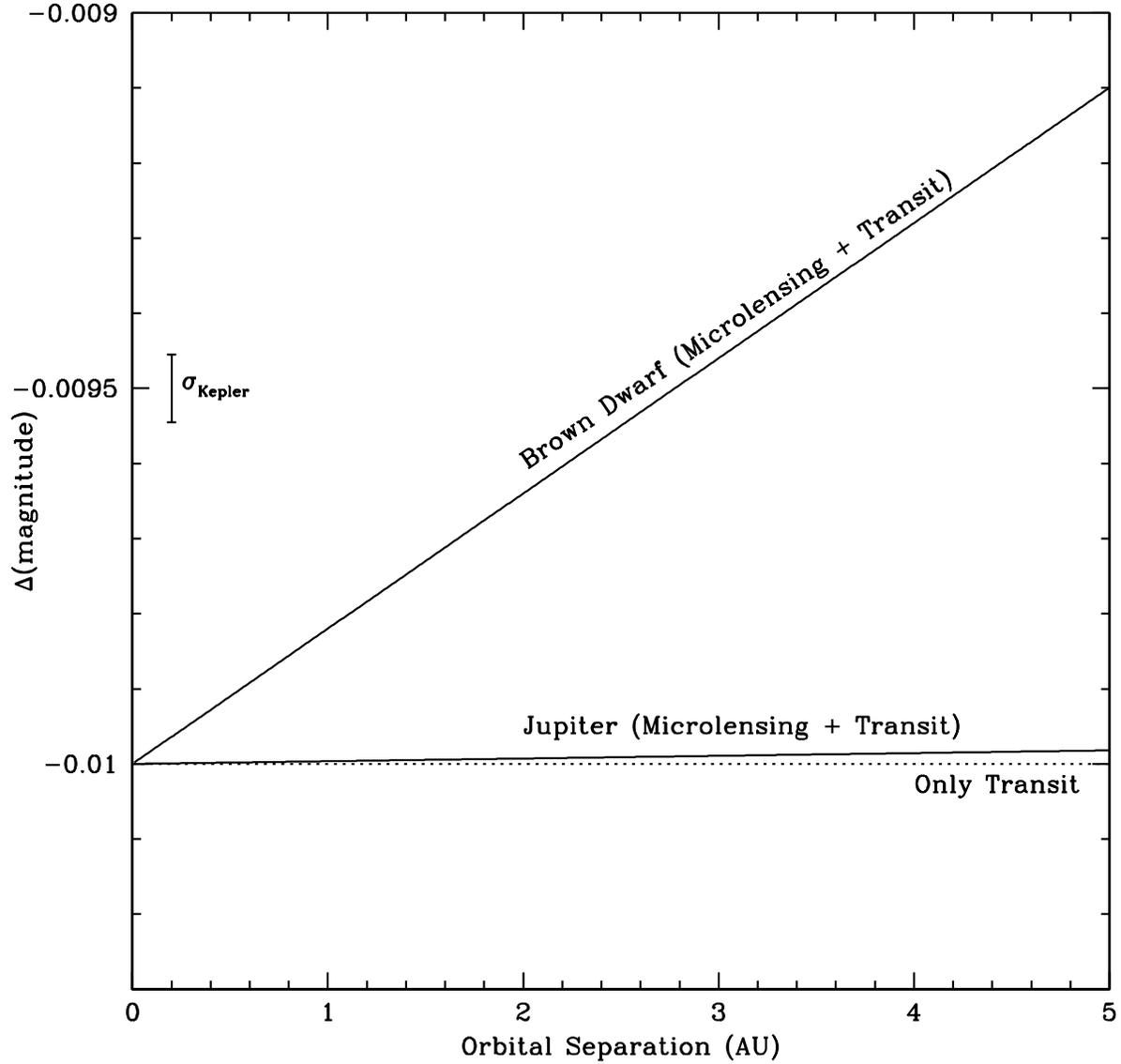}
\caption{Same as the previous figure  but the y-scale is enlarged to show in detail
the effects of a brown dwarf and a Jupiter-mass planet. At an orbital
separation of 1 AU, the   the effect of microlensing for a Jupiter-mass
planet is negligible. But the effect of microlensing by a brown dwarf
is appreciable and must be taken into account in deriving its physical
parameters from the photometric observations. At an orbital separation
of 5 AU, the effect of microlensing is just noticeable for a
Jupiter-mass planet.}
\end{figure}

\begin{figure}
\plotone{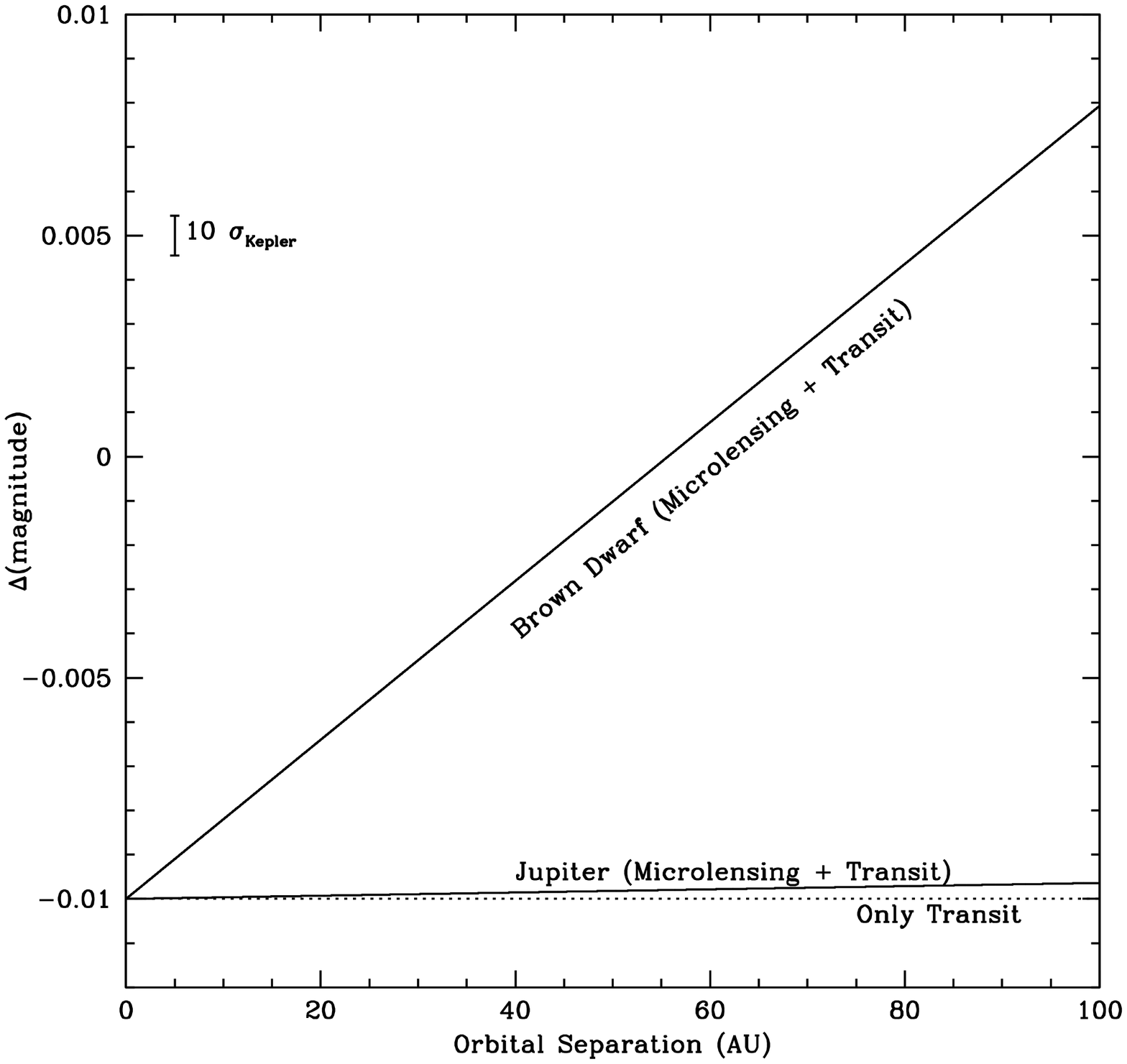}
\caption{Same as the previous figure but for a larger range of orbital separations
(0 to 100 AU). This figure shows that for a brown dwarf secondary, the
transit signal is fully  compensated by the microlensing signal at an
orbital separation of $\sim$55 AU, after which the net effect is a
positive amplification.}
\end{figure}

An enlarged view of the effects of the white dwarf, neutron star and
black hole are shown in Fig. 8 for small orbital separations of 0 to
0.1 AU. For neutron  star and black hole secondaries, the net
amplification is positive at any orbital separation. For a white dwarf,
the microlensing contribution almost  exactly cancels out the transit
contribution at $\sim$0.05 AU, beyond  which the net amplification is
positive. 

\begin{figure}
\plotone{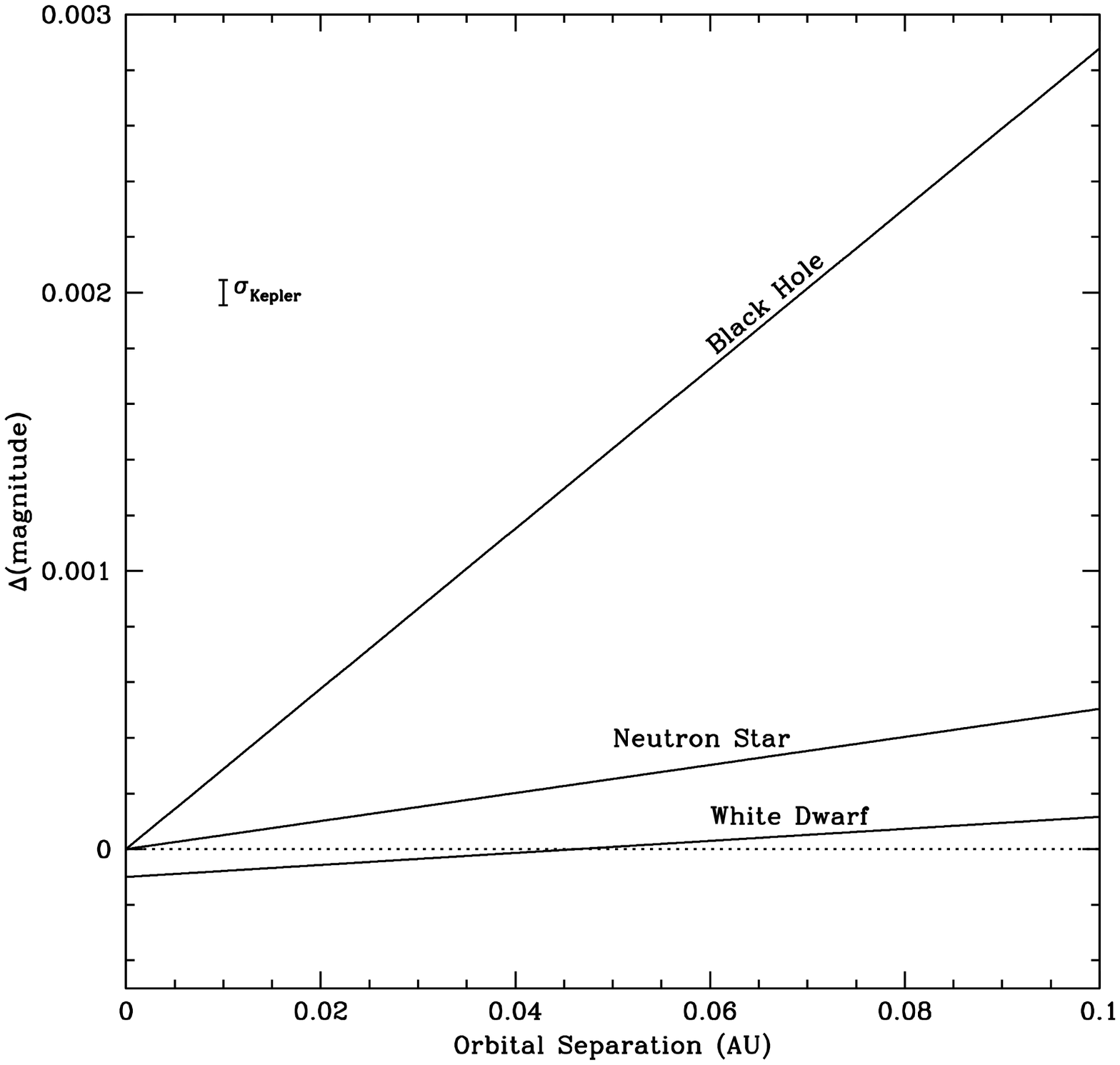}
\caption{The amplifications caused by white dwarf, neutron star and
black hole secondaries at small orbital separations (0 to 0.1 AU). In 
case of the black hole and  neutron star secondaries, the microlensing
contribution is larger than the  transit signal even for the smallest
possible orbital radius (when the orbital separation is the same as the
stellar radius), so that the net  amplification is always positive. For
a white dwarf  secondary, the net amplification is positive at an
orbital separation of 0.046 AU and beyond.}
\end{figure}

The effect of microlensing is more dramatic if the size of the source
is smaller, as is the case for a brown dwarf. Fig. 9 shows such a case
where we assume the size of the source to be $7 \times 10^{9}$ cm, as
applicable for a brown dwarf or a Jupiter. The amplification, which is
the ratio of the amplified luminosity and the original luminosity, is
typically much larger compared to the earlier case of a solar type
source. 

\begin{figure} 
\plotone{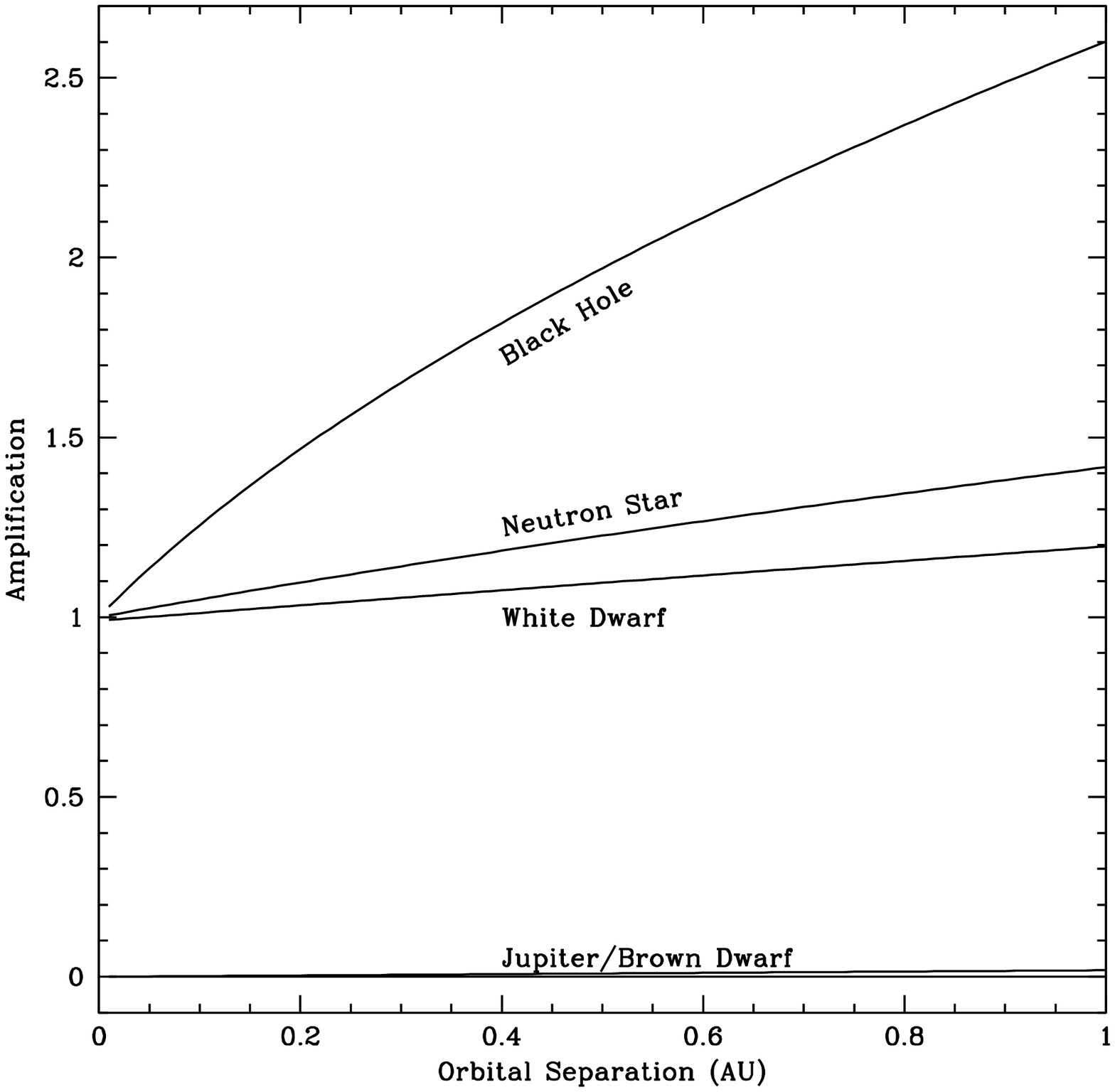} 
\caption{The effect of microlensing when the size of the source is
$7 \times 10^{9}$ cm, as applicable for a brown dwarf or a Jupiter.
The effect is more dramatic when the source size is smaller.  The
amplification, which is the ratio of the amplified luminosity and the
original luminosity, is much larger compared to the earlier case of a
solar type source.}
\end{figure}

The expected amplification is a function of inclination, which is shown
in Fig. 10. The solid curve is the expected  maximum transit depth, and
the dotted curve shows the expected amplification at mid point taking
both the transit and the microlensing into account. Here the source
(primary) is assumed to be a solar type star and lens (secondary)  is a
white dwarf at an orbital separation of 1 AU.

\begin{figure}
\plotone{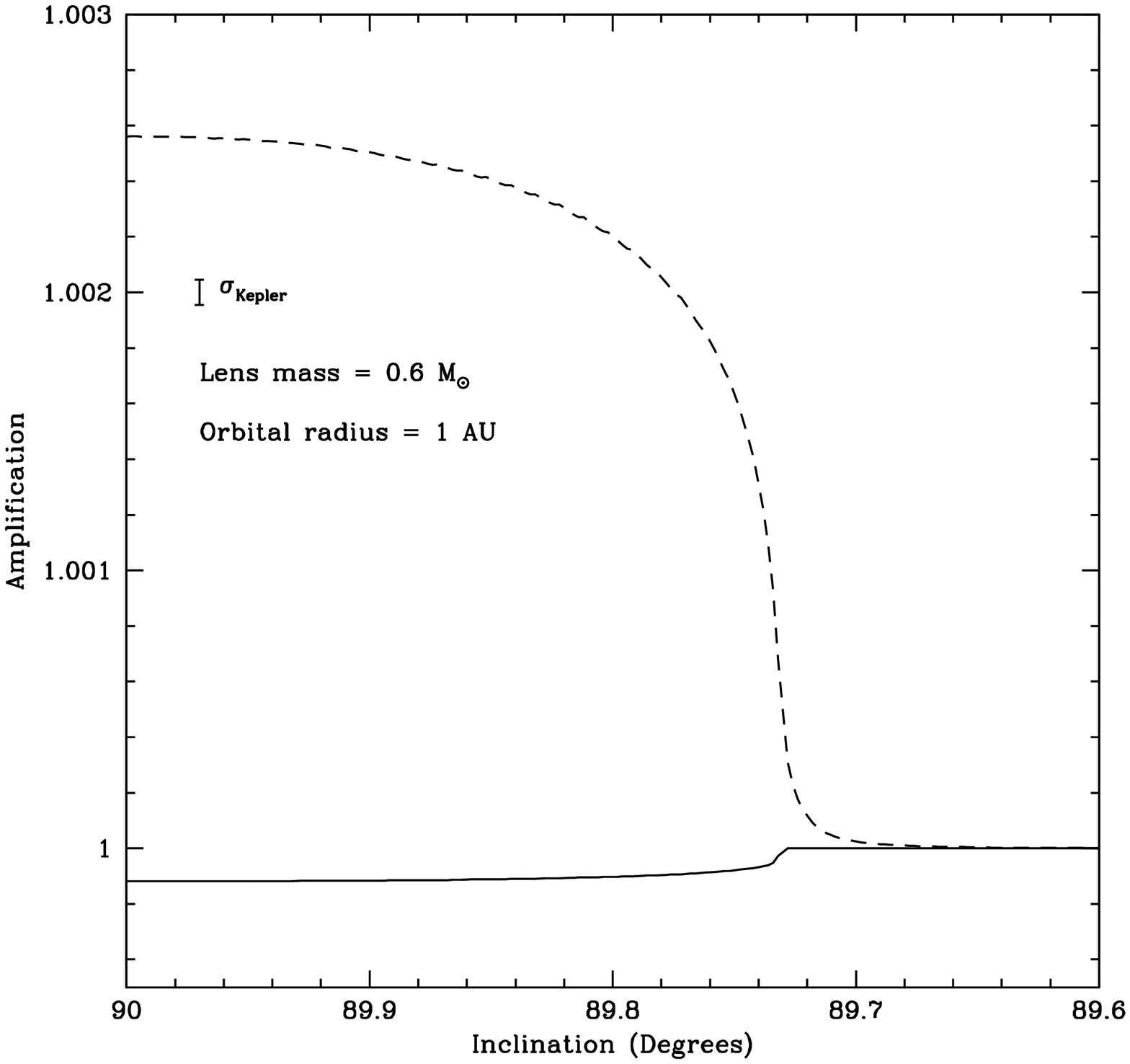}
\caption{The expected maximum amplification caused by a white dwarf at
1 AU from a solar type star is shown here as a function of inclination. 
The solid curve is the pure transit contribution, and the dashed curve
takes both the microlensing and the transit contributions into account.
The limb darkening of the source is taken into account in generating
the curves.}
\end{figure}

To show the relative importance of microlensing and transit for various
kinds of secondaries at different orbital separations, we have plotted
in Fig. 11 the orbital radius at which the microlensing signal equals
the transit signal as a function of the mass of the secondary. For
example, for a white dwarf secondary, the expected transit depth, given
by $r_L^2/r_S^2$, where $r_L$ and $r_S$ are the  radii of the secondary
(lens) and the primary (source) respectively, is 0.0001 and the  mass
of the lens is 0.6 M$_\odot$. The corresponding orbital separation at 
which the microlensing signal equals the transit signal is 0.046 AU.
For an   Earth-mass object, the corresponding orbital radius is
$\sim$2300 AU.

\begin{figure}
\plotone{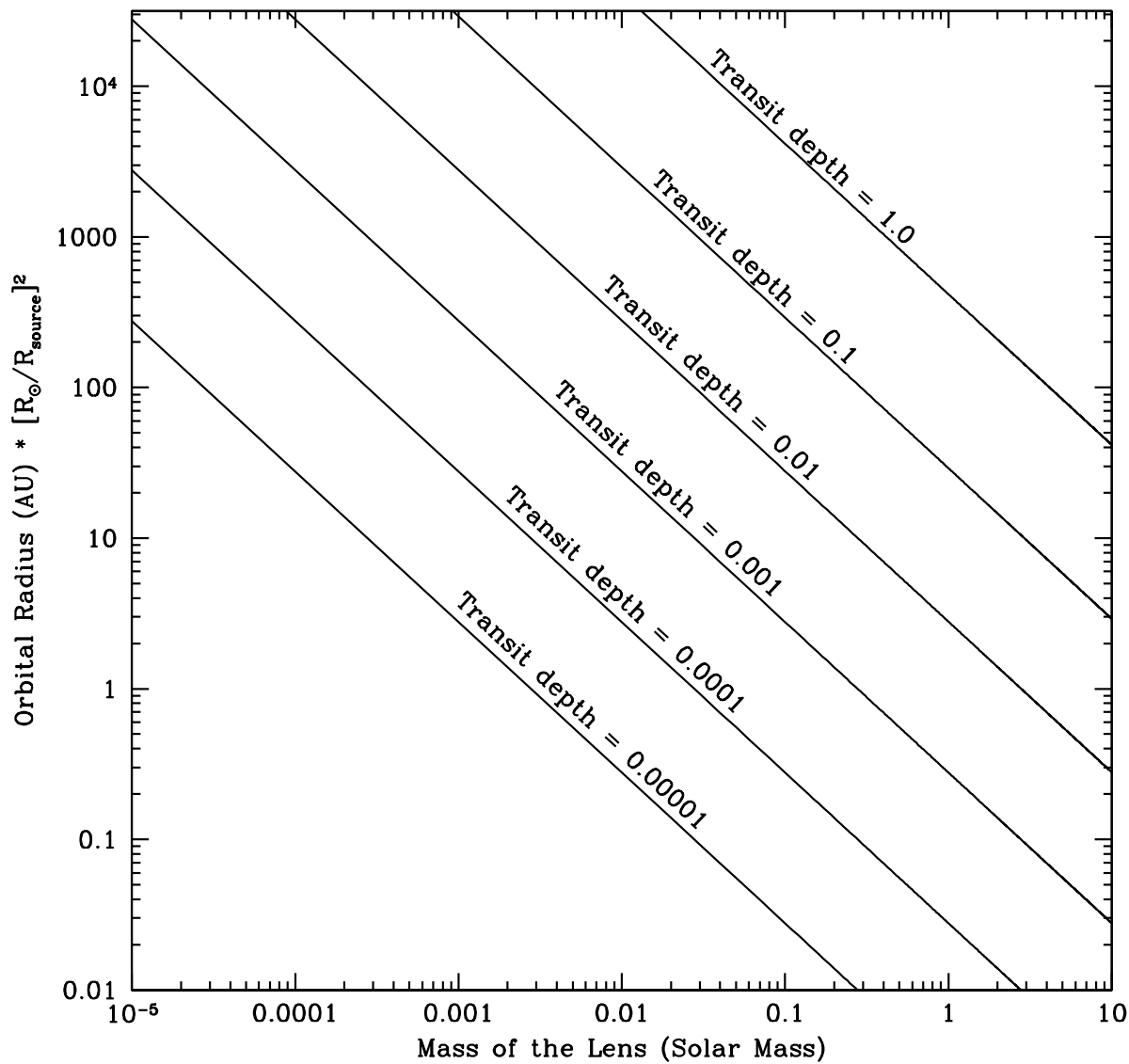}
\caption{The orbital radius at which the microlensing signal is
equal to the transit signal is plotted here as a function of the mass
of the secondary, for various transit depths. For example, for a white
dwarf, the expected transit depth is 0.0001 and the mass of the
secondary is $\sim$0.6 M$_\odot$. The corresponding orbital radius is
0.05 AU where the transit signal equals the microlensing signal, and
beyond which the net amplification is positive.}
\end{figure}

\section{Special Cases}

\subsection{Jupiters and Brown Dwarfs}

The radii of brown dwarfs are similar to the radii of Jupiters, so the
expected light curves due to their transits would be similar. Thus, a
brown dwarf is indistinguishable from a Jupiter from its transit signal
alone. However, since the mass of a brown dwarf is considerably larger 
(15-80 M$_{Jup}$), the microlensing signal from a brown dwarf would be
much larger. For example, Fig. 12 shows the light curves as expected
from a Jupiter and a 50 M$_{Jup}$ brown dwarf at 1 AU. The dashed curve
shows the pure transit for both objects and the solid curve shows the
combined effect of the transit and microlensing for a brown dwarf. For
a Jupiter, the combined effect of transit and microlensing is
indistinguishable from a pure transit curve. For a brown dwarf, the
combined light curve is different from the transit curve. Note that the
expected depth in the observed light curve is smaller for a brown dwarf
than for a Jupiter, but the shapes of the light curves for both objects
are very similar. Thus, there is a degeneracy between the light curves
due to a Jupiter and a brown dwarf. It is clear that the contribution
due to microlensing must be taken into account in the interpretation of
the observed light curves, particularly in estimating the mass/radius
of the secondary. Fig. 13 shows the light curves  for a brown dwarf at
55 AU which, if observed, would be easy to interpret as due to a brown
dwarf since the light curve due to a  Jupiter is close to the transit
curve itself. Notice that there is a slight positive amplification at
the ingress (egress) since the microlensing begins before (ends after)
the transit contribution begins (ends). This curve corresponds to an
orbital period of 405 yr, and is hence mostly of pedagogical interest
since the probability of observing such an event is very small.

\begin{figure} 
\plotone{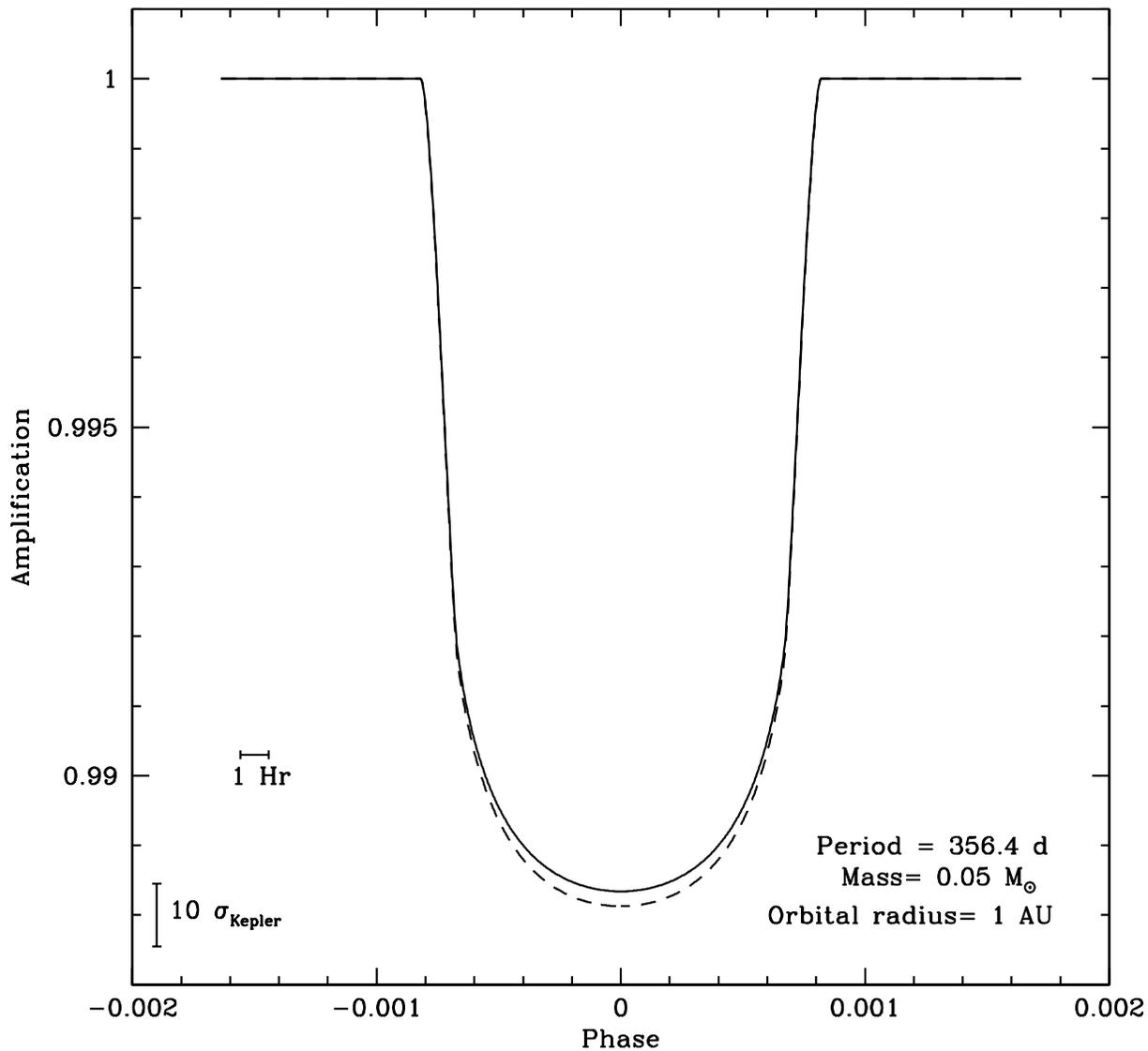} 
\caption{The light curves caused by a 0.05 M$_\odot$ brown dwarf 
and a Jupiter at 1 AU from a  solar type star. The dashed curve is what
one would expect purely from transit. The solid curve shows the
combined effect of the transit and microlensing for a brown dwarf.
The combined effect of transit and microlensing is indistinguishable  
from a pure transit curve for Jupiter since the shape of the
combined as well as the transit light curves are very similar. The
depth of the observed light curve would be smaller  for a brown   
dwarf than for a Jupiter. }
\end{figure}

\begin{figure} 
\plotone{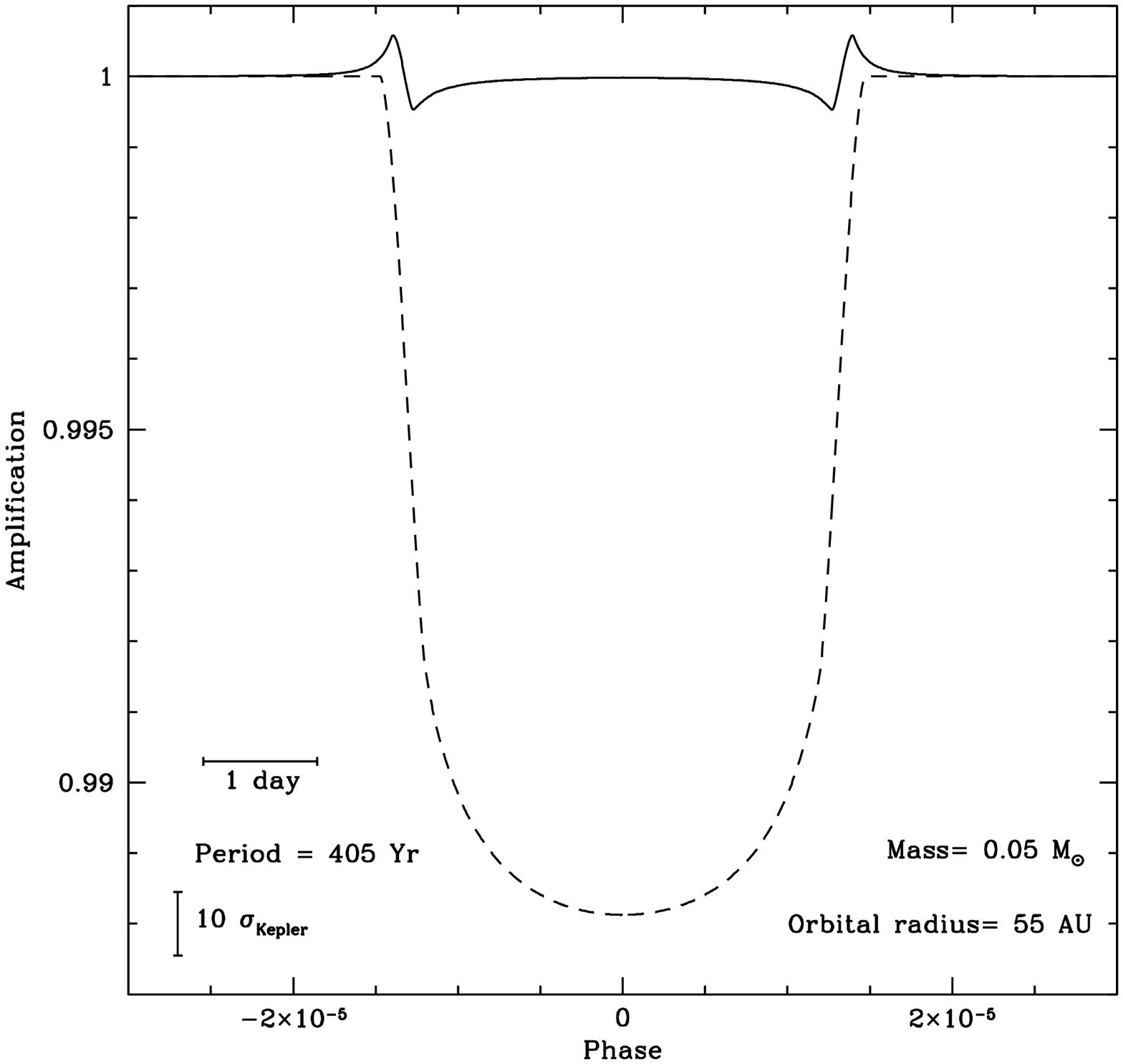} 
\caption{The light curves caused by a 0.05 M$_\odot$ brown dwarf at
55 AU from a  solar type star. The dashed curve is what one would
expect purely from transit. The solid curve takes both the transit and
the microlensing contributions into account, and hence corresponds to
the actual light curve that one would observe.  Notice that there is a 
slight positive amplification at the ingress (and egress) since the
microlensing begins before (ends after) the transit contribution begins
(ends). This curve corresponds to an orbital period of 405 yr (and is
hence mostly of pedagogical interest).}    
\end{figure}

\subsection{Earths and White Dwarfs}

Interestingly, white dwarfs have radii similar to Earth-mass planets.
As mentioned earlier, one of the prime objectives of {\em Kepler} is to
detect Earth-mass planets from their transit signals.  There has been
some concern that the Earth-mass planets would be confused with white
dwarfs since their transit signals would be similar. (The secondary
eclipse may provide a clear test to distinguish between a white dwarf
and an Earth-mass planet, but the secondary eclipse may not be seen in
some cases when the orbital eccentricity is high or because  the
timescale is such that only the primary eclipse is observed). However,
as we explained in \S 4, in the case of a white dwarf the  effect of
microlensing will dominate the signal. Fig. 14 shows the light curves
due to a white dwarf secondary at 0.03 AU, 0.0463 AU and 0.1 AU, with
corresponding periods of 1.50, 2.88 and 9.13 days, respectively. The
dashed curves are the  light curves expected from pure transit, and the
dotted, solid and dot-dashed curves are the corresponding light curves
after taking the microlensing contribution into account. Only at
orbital separations smaller than 0.0463 AU would the expected
$\Delta$mag be negative, with the shape of the observed light curve
similar to the transit curves.  We note that at an orbital separation
of less than 0.01 AU, the primary is likely to fill its Roche lobe
(Eggleton, 1983). This will cause mass-transfer from the primary, which
may cause light variations  much larger than the transit or the
microlensing effect. Also, the primary will be tidally distorted
producing distinct ellipsoidal variations. Thus, the transit effect 
may be difficult to observe for separations below  0.05 AU, beyond
which the $\Delta$mag is expected to be positive.

\begin{figure} 
\plotone{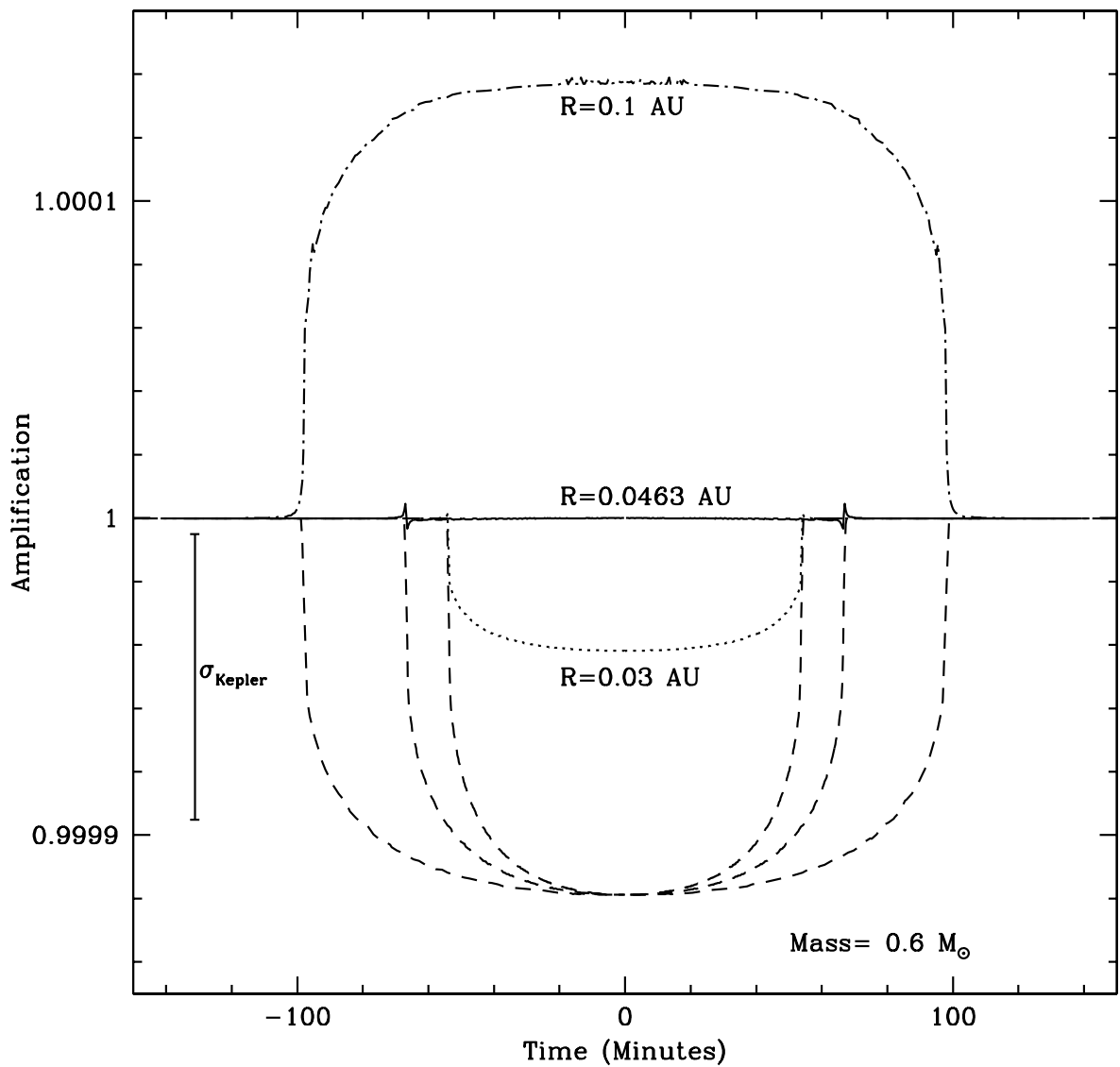} 
\caption{The light curve caused by a 0.6 M$_\odot$ white dwarf at
0.03, 0.0463 and 0.1 AU from a  solar type star. The dashed curves are
what one would expect purely from transit. The corresponding upper
curves (dotted, solid and dot-dashed curves, respectively) take both
the transit and the microlensing contributions into account, and hence
correspond to the actual light curves that one would observe. Notice
that, even at a small orbital radius of 0.03 AU (where the white dwarf
is likely to produce tidal distortions on the companion), the
microlensing contribution is significant. At an orbital radius of
0.0463 AU, the microlensing contribution almost exactly cancels out the
transit contribution. At an orbital radius of 0.1 AU, the net effect is
a positive amplification which is easily detectable by {\it Kepler}.}
\end{figure}

The expected light curve when the white dwarf is at 1 AU is shown in
Fig. 15. In such a case, the white dwarf would be detectable through
its positive amplification at $>$100 $\sigma$-level in
observations averaged over a single event with {\em Kepler}!

\begin{figure}
\plotone{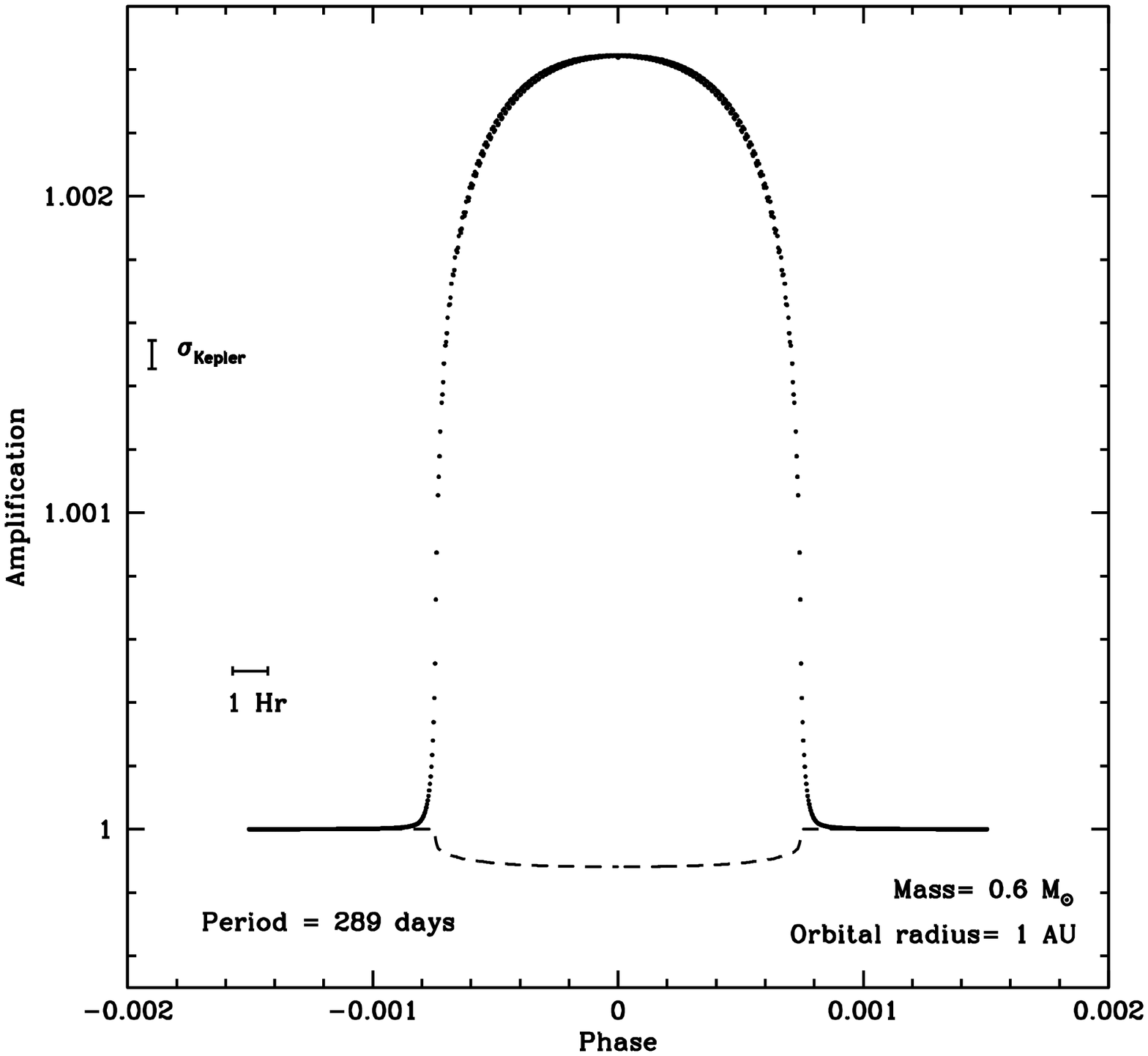}
\caption{The light curve caused by a 0.6 M$_\odot$ white dwarf at 1
AU from a  solar type star. The dashed curve is what one would expect
purely from transit. The solid curve takes both the transit and the
microlensing contributions into account, and hence corresponds to the
actual light curve that one would observe. Notice that the white dwarf
would be detectable through its positive amplification at $>100$
$\sigma$-level in observations averaged over a single event with {\em
Kepler}!}
\end{figure}

It is of interest to consider how accurately {\em Kepler} observations
of a White Dwarf transiting a solar-like star in a 1 AU orbit would
constrain the lens mass.  For such a  canonical case (shown in Fig.
15), the maximum amplification is 0.0024 magnitudes.  The precision per
15 minute sample for an assumed $V$ = 12 star observed with {\em
Kepler} is 90 $\mu$mag. Hence, with 40 samples during the 10 hour
lensing event, the maximum amplification will be determined to about 15
$\mu$mag, resulting in a relative error on $A_{0}$ of 0.6\%. Given that
the mass of the lens scales as the square of the amplification
(equations  15-16), and assuming that four events are available for
averaging,  the relative error contribution for the lens mass from the
microlensing observations is also 0.6\%.  However, equations (15) and
(16) also contain the mass and radius of the source star for which we
do not have direct values available from the {\em Kepler} observations
alone.  Radial velocity observations of even moderate quality would
provide a further  constraint on the mass, leaving uncertainties on the
radius of the  source star, $r_S$, as the likely largest contributing
error source. Equation (17) provides a direct constraint on $r_S$
simply from the event width, but the unknown orbital inclination must
also be allowed for. Spectroscopic measurements might suffice to
constrain $r_S$ at about  10\% level, thus the lens mass to 20\% level.
 As was done for HD 209458 (Brown {\em et al.} 2001), constraints on
the source  stellar radius may also be derived from the light curve
itself, since some subtle details will depend in unique ways on the
generally degenerate effect of $r_S$ and $M_L$ in fixing the light
curve.  Equation (21) provides a further constraint on the lens mass if
an estimate of the mass of the primary is available. Note that the
orbital  inclination and limb darkening parameters can also be
estimated from the observed light curve (as was done for HD 209458), 
and a detailed treatment of the errors would be required in determining
the uncertainties in different physical parameters. Fitting the  entire
light curve would take all the constraints in a self-consistent manner,
providing the best determination of the various parameters, including
the lens mass to an estimated accuracy of 1 to 5 percent. 

\subsection{Neutron Stars and Black Holes}

The expected transit signals from a neutron star and a black hole are
of the order of nano-magnitudes and smaller. So the transit signal in
these cases can be neglected for all practical purposes. In the
treatment of microlensing, we note that the deflection corresponding to
the fainter image (i.e. the smaller of the two deflections) is larger
than the size of the lens even for small orbital separations. So the
finite size of the lens can also be ignored. Thus the expected light
curve is the same as what is expected from the pure microlensing of an
extended source and a point lens, e.g. as in Fig. 15 with amplitude
scaled approximately linearly by the lens mass.  Note that the expected
$\Delta$mag from a neutron star and a black hole at an orbital radius
of 1 AU are 5 and 28 milli-magnitudes respectively, which may be
detectable from ground-based observations. 

\section{Discussion and Summary}

We have shown that near-field microlensing of main sequence stars by
orbiting compact companions provides signals that would be easily
detected by a space mission such as {\em Kepler} dedicated to the
detection of  extrasolar terrestrial planets via transits.  For
terrestrial planets with orbital periods up to a few years to which
{\em Kepler} has sensitivity to transits, the microlensing contribution
to the signal will always be entirely negligible.  Jovian class planets
in orbits of a few AU provide a microlensing contribution that just
starts to be relevant, albeit still small compared to the inherent
transit signal.  White dwarfs, neutron stars, or black holes transiting
a main sequence star in an $\sim$1 AU orbit would provide easily
detectable  microlensing signatures that swamp any underlying transit
signal. Therefore, although the radius of a white dwarf is similar to
that of a terrestrial planet, a transiting white dwarf (with an orbital
separation of 0.25 - 2.0 AU, which is the region of prime interest for
{\em Kepler}) will not provide a source of false alarms for extrasolar
planet detection since the resulting signal will be a  significant
brightening rather than a small diminution in light.
 
Compact objects like white dwarfs, neutron stars, and black holes are
now easily found when they reside in close orbits ($\lesssim$ 0.1 AU) 
from a main-sequence companion since they induce ellipsoidal variations
in the companion. Furthermore, such systems often become x-ray and
radio sources through emissions associated with mass transfer (even
when the mass transfer occurs at low levels). While the microlensing
signal may be detectable even for an orbital separation as small as
$\sim$0.1 AU,  other variations (such as ellipsoidal variations caused
by the distortion of the companion, variations from star spots for
induced rapid  rotation) clearly call attention to the presence of a
degenerate object in such systems.  As an example, we note that KPD
1930+2752 is one such interesting  case of a sdB binary in which the
unseen companion is likely to be a white dwarf with an orbital period
of 0.095 days (Maxted et al. 2000, Billeres et al. 2000). 

The microlensing signal for compact objects with orbital separations of
$\gtrsim$1 AU will be easily seen by {\em Kepler}. In such cases, old
white dwarfs, neutron stars and black  holes would not already have
signaled their presence since the wide separation would not induce the
canonical signatures discussed above.  However, as with  extrasolar
planet transits, microlensing from orbiting compact objects will only
be observed in a very small fraction of cases in which the 
inclinations happen to be ideal so that transits can be observed.  The
probability of transits is $r_S /a$, where $r_S$ is the radius of the
primary, and $a$ is the orbital separation.  For a solar like star and
an object at 1 AU this probability is about 0.005.  Thus if all 100,000
stars in the {\em Kepler} sample have compact objects orbiting within
0.25 to 2.0 AU, some 500 would be detected through their microlensing
signatures.  If a more physically reasonable  fraction of 1\% of stars
surveyed contain such compact, massive objects then the number detected
by {\em Kepler} would be 5. What, then, are the prospects for observing
compact objects via microlensing signatures by {\em Kepler}?  Duquennoy
\& Mayor (1991) and Heacox \& Gathright (1994) show that about
two-thirds of G-type stars are binaries, and that about 20\% of them
have periods between 0.25 and 2.0 years.  Thus, for the {\em Kepler} 
mission to detect a statistically modest total of 5 systems with 
microlensing in such circumstances, $\sim$7.5\% of binary systems must
contain a compact object (most likely a white dwarf). By comparison,
the observed white dwarf fraction in the solar-age open cluster M67 is
about 9\% by mass \cite{richer98}, and since incompleteness may still
be important in this case, this is a lower limit to the detected
cluster white dwarfs. This suggests that the overall fraction of white
dwarfs is sufficient to yield several  detections during the {\em
Kepler} mission.  Estimates of the fraction of baryonic mass in compact
objects in the local disk typically span  values of 10 -- 30\% (e.g.
Weidemann, 1990).  This, in turn, implies that the detected number of
such microlensing systems will be able to place interesting constraints
on the fraction of stars having compact companions.

For microlensing detections from systems where the orbital period is
known from repeated events, it will be possible to determine the mass
of the compact object directly from the amplitude of the microlensing
signal (assuming independent information on the primary main sequence
star is available).  For the case of white dwarfs, information on the
compact object may also be available if the secondary eclipse is
observed, and radial velocity followup observations will be able to
determine independent mass estimates.

\acknowledgements We thank Meena Sahu and Mario Livio for useful
discussions. We wish to thank the referee, Scott Gaudi, for useful
suggestions.

\end{document}